\documentclass[article]{elsarticle}

\newcommand{\mathbb}[1]{\mathbf{#1}}

\usepackage{media9}
\usepackage{multimedia}
\usepackage{media9}

\usepackage{color}

\begin{document}

\title{Spreading processes in post-epidemic environments}

\author[a,b]{V. Blavatska\corref{cor}}\ead{blavatskav@gmail.com}
\author[a,b,c]{Yu. Holovatch}
\cortext[cor]{Corresponding author}
\address[a]{Institute for Condensed Matter Physics of National Acad. Sci. of Ukraine, Lviv, Ukraine}
\address[b]{L$^4$ Collaboration \& Doctoral College for the Statistical Physics of Complex
Systems, \\Leipzig-Lorraine-Lviv-Coventry, Europe}
\address[c]{Centre for Fluid and Complex Systems, Coventry University,\\ Coventry, CV1 5FB,
United Kingdom}

\date{\today}

\begin{abstract}
We analyze infection spreading processes in a system where only a fraction $p$ of individuals can be affected by disease, while remaining
	$1-p$ individuals are immune. Such a picture can emerge as  a natural  consequence of  previously terminated epidemic process or arise
	in formerly vaccinated population. To this end, we apply
	 the synchronous cellular automata algorithm studying stationary states and spatial patterning in SI, SIS and    SIR models on
	a square lattice with the fraction $p$ of active sites. A concept of  ``safety patterns'' of susceptible agents surrounded by immune individuals
	naturally arises in a proposed system, which plays an important role in the course of epidemic processes under consideration.
	Detailed analysis of distribution of such patterns   is given, which in turn determine the fraction of infected agents in a stationary state $I^*(p)$.
	Estimates for the threshold values of the basic reproduction number $R_0^c$ as a function of active agents fraction $p$ are obtained as well.
	 In particular, our results allow to predict the optimal fraction of individuals,  needed to
          be vaccinated in advance in  order to get the maximal values of  unaffected agents  in a course of epidemic process with a given curing rate.

\end{abstract}


\maketitle

\section{Introduction}

The frequency and magnitude of epidemic processes is observed to increase  in the last decades
mainly due to globalization, growth of large centers of population in industrialized society and thus the
significant growth of social contacts.
Mathematical modeling in epidemiology allows to explain the dynamics patterns of epidemic process course
and to propose mechanisms of disease control
by providing a deeper understanding of  factors controlling disease persistence  \cite{Brauer,Bailey}.

The basic concepts of epidemics modeling were established in pioneering work of Kermack and
McKendrick \cite{Kermack27}.
The population is assumed to be homogeneously mixed and contains three main classes  {-- compartments --} of individuals:
susceptible $S$ (which are healthy and can be infected),
infected $I$ (able to transmit the disease via air or physical contact), and removed $R$ (either by recovery
and acquiring immunity or by death).
The way each particular infection is spread and cured is described by simple differential equations governed by  parameters $\beta$ and $\gamma$ \cite{SIAM}.
The {\it infection rate} $\beta$ is dependent on the mechanism of infection transmission
(air or direct physical contact) and may in general be tuned by changing the averaged frequency of contacts of individuals (e.g. it is sensitive to seasonal
fluctuations  \cite{Dietz76,Yorke79} or age of individuals \cite{Anderson85a}).
 The {\it curing rate} $\gamma$ is basically intrinsic for the given disease and its progression.
Their ratio $R_0=\beta / \gamma$  called {\it basic reproductive number}  is the expected
number of secondary cases produced
by an infectious individual in a population \cite{Dietz76,Milligan15}
and is
directly measured for a number of diseases.


Three principally different scenarios, directly observed in epidemiology, are perfectly captured within the frames of three corresponding
 simplest models called SI, SIS and    SIR.
Within the frames of SI model,  individuals without proper treatment stay infected and infectious throughout their life and
remain in contact with the susceptible population. This model matches the behavior of diseases like cytomegalovirus (CMV) or herpes.
The SIS model describes the disease with no immunity, where the cured individuals can be infected repeatedly.
This model is appropriate for diseases like the common cold (rhinoviruses). It consequently describes the emergence of so-called {\it endemic state}
of the disease, where starting with $R_0>1$ the infection spreading process continues in time without termination.
The    SIR model describes the opposite situation when recovered person becomes immune, as in the case of influenza or measles.
This mechanism leads to possibility of epidemic  outbreak (at $R_0>1$).
Further refinements have been done by taking into account the latency period (with its
duration dependent on the type of the disease, the dose of infectious agent, the host immune response etc.)
by introducing the additional compartment for so-called exposed individuals (E). This leads
to so-called SEIS and SEIR models. In particular,    SIR and SEIR models
have been used to describe and predict epidemic processes of respiratory infections (influenza A, H1N1 etc.)
 \cite{Beauchemin05,Dorjee,Coburn09,Kumar16,Lopez17} as well as COVID-19 \cite{Godio20,Arino20,Neufeld20,Kaxiras20,Pan20,Feng21,Lux21,Faranda20,Burda20,Frenkel21}.

In the classical models of infection spreading  it is assumed, that at initial
stage the population is homogeneous and individuals are equally susceptible to the disease.
However, heterogeneity in susceptibility and infection is an important feature
of many infectious illnesses. The impact of variable infection rates, which are
heterogeneously distributed in population, on epidemic growth have been analyzed in Refs. \cite{Andersson98,Muller07,Rodrigues09,Corral}.

Another approach of mathematical modeling of disease spreading is based on discrete lattice models,
motivated by cellular automaton implementation of synchronous-update Markovian process
on a lattice  \cite{Grassberger83,White07}.
In particular, such approach  allows to take into account the local characteristics of the system, e.g. to  include variable susceptibility
 for different individuals, and to analyze  the  patterns (clusters) formation in spreading process \cite{Hiebeler,Corral, Ilnytsky16,Ilnytsky18}.
The evolution of system is discrete in time and is governed by a specific update algorithm which
 takes into account the state of each agent \cite{White07,Fuentes99,Ahmed98,Beauchemin05}.
In particular,  the special case when  $\beta=1-\gamma$ in SIS cellular automata model (with $R_0=(1-\gamma)/\gamma$),
 can be directly connected to dynamics of the so-called contact process
 \cite{Griffeath,Sabag}, which exhibits   {a non-equilibrium} phase transition with critical behavior
 of the universality class of directed percolation  {\cite{Odor04}}.
 The critical value of creation rate $\lambda_c$ of this model thus is directly related with the threshold value of reproduction number $R_0^c$ in
 disease spreading process.
 In contrast to continuous SIS  model, in this case the threshold value of reproduction number which separates the non-spreading and
 endemic states is $R_0^c=1.64874(4)$ \cite{Sabag}, and thus is  larger than   {the result obtained within homogeneous
 mixing assumption $R_0^c=1$}.
 Corresponding critical value of curing rate $\gamma_c=0.3775$.
In the    SIR cellular automaton, it was argued \cite{Cardy85}  that  a continuous phase transition  between non-spreading and spreading regimes is
of the bond percolation universality class.
{ A deep connection between the epidemy spreading  and percolation theory is  discussed in Ref. \cite{Ziff21}.}

Again for the special case when  $\beta=1-\gamma$ one obtains the critical curing rate $\gamma_c=0.1765005(10)$
and corresponding threshold value for reproduction number $R_0^c=4.665706$ \cite{Ziff10}.
It thus follows that the epidemic spreading for the SIS model occurs
at a smaller value of the reproduction number  comparing with the    SIR model. This is intuitively clear since
  {the susceptible individuals in the SIS model} can be infected multiple times.


In the present work, we consider the epidemic spreading on the lattice model of population
with the fraction $p$ of individuals susceptible for a particular disease, whereas the rest $1-p$ randomly chosen agents
 are considered to pertain immunity to it. On the one hand, such situation can describe  {an existence of} the congenital
 immunity or vaccination performed against the given disease.
On the other hand, it may correspond to the starting point of the "second wave" of an epidemic process,
when some fraction of population already get removed (or obtained  lifelong immunity) after termination of the first wave.
The proposed model is related to the contact processes on heterogeneous lattices, where
the spreading is restricted only to the fraction $p$ of available sites \cite{Moreira96}. There, the threshold value of the
reproduction number  $R_0^c$ was shown to increase when $p$ decreases.
The related problems of contact processes on heterogeneous surfaces in the shape of deterministic fractals like Sierpinski gasket
\cite {Mai92,Tretyakov91,Jensen91,Lee08} or two-dimensional percolation cluster \cite{Tretyakov91,Casties93} have also been studied.
  We will use the cellular automata implementation of the three basic epidemic models SI, SIS and    SIR on a
two-dimensional regular lattice with variable ratio $p$ ($0\leq p \leq 1$) of sites containing agents involved in epidemic process.

The layout of the paper is as follows. In the next Section, we give the scheme of cellular automata algorithm applied for modeling the spreading processes.
In Section \ref{III}, the peculiarities of cluster size distributions on disordered lattices are discussed. Section \ref{IV} contains our main results
 {for description of SI, SIS and    SIR spreading scenarios on spatially-inhomogeneous lattices}. We end up with Conclusions in Section \ref{V}.

\section{Cellular automata algorithm of modeling epidemic processes} \label{II}

Let $S(t)$, $I(t)$ and $R(t)$ be fractions  {at time $t$} of healthy (susceptible), infected and recovered individuals,   {and}
$S(t)+I(t)+R(t)=1$. Parameters
$\beta$ and $\gamma$ are correspondingly infection and curing rate.
We consider the population of individuals (agents), located  on the sites of a
regular two-dimensional  {square}  lattice of size $L\times L$. The links between each site and its four nearest neighbours
denote the possible local contacts of a corresponding individual. At time $t$, the $k$th site of the lattice
is in a state $\sigma_k(t)$, with $\sigma_k(t)$ taking values from a set  $\{0,1,2\}$,
corresponding to {\em S}, {\em I}, or {\em R}  {respectively}.
Thus, the global fractions  {\em S}, {\em I}, and {\em R} are given by:
\begin{eqnarray}
&&S(t)=\frac{1}{N}\sum_{k=1}^N\delta_{0\sigma_k(t)},\nonumber\\
 &&I(t)=\frac{1}{N}\sum_{k=1}^N\delta_{1\sigma_k(t)},\nonumber\\
 && R(t)=\frac{1}{N}\sum_{k=1}^N\delta_{2\sigma_k(t)},\nonumber
\end{eqnarray}
where $\delta$ is the Kronecker delta and $N=L\times L$ is total number of agents.
System evolution is governed by an  update algorithm which changes the state of a particular individual according
to the states of its neighbors.  We will consider the  synchronous process,  {where one time step
implies a sweep throughout the whole lattice \cite{Grassberger83}.  Time update $t-1 \to t$ consecutively considers all $k=1, \dots, N$
lattice sites and changes each state $\sigma_k(t-1) \to \sigma_k(t)$ according to the spreading scenario rules.
The updated state of each lattice site is stored in a separate array, and after the sweep is finished,
the old state is replaced by the updated one. Update rules for the three spreading
scenarios considered here read:}

{\em (i) SI model}
\begin{itemize}
	\item  {Choose site $k$}
	\item If $\sigma_k(t-1)=0$,  {then do nothing}	
\item  If $\sigma_k(t-1)=1$, then one of its randomly chosen susceptible neighbors $j$ (with $\sigma_j(t-1)=0$)
changes its state to 1 with probability $\beta$, so that $\sigma_j(t)=1$.
\end{itemize}
{\em (ii) SIS model}
\begin{itemize}
	\item  {Choose site $k$}
	\item  {If $\sigma_k(t-1)=0$, then do nothing}	
  \item If $\sigma_k(t-1)=1$, then with probability $\gamma$ it is cured (so that $\sigma_k(t)=0$), otherwise
 one of its randomly chosen susceptible neighbors $j$ changes its state to 1 with probability $1-\gamma$, so that $\sigma_j(t)=1$.
\end{itemize}
{\em (iii) SIR model}
\begin{itemize}
	\item  {Choose site $k$}
	\item  {If $\sigma_k(t-1)=0$ or $\sigma_k(t-1)=2$  then do nothing}	
  \item If $\sigma_k(t-1)=1$, than with probability $\gamma$ it is recovered (so that $\sigma_k(t)=2$),
otherwise  one of its randomly chosen susceptible neighbors $j$ changes its state to 1 with probability $1-\gamma$, so that $\sigma_j(t)=1$.
\end{itemize}
We consider a  square lattice of size $L=300$, the total number of individuals thus being $N=90 000$.
 The periodic boundary conditions are applied in all directions.
As already mentioned above, we will analyze spreading processes on a lattice, where only a randomly selected part of
 sites (agents) is susceptible to disease. To this end, we consider each  randomly chosen site $k$ of the lattice to be either susceptible to disease with
 probability $p$
or immune with  probability $1-p$ (the state of corresponding sites  $\sigma_k=0$ always, and these sites are considered as
 non-active in updating algorithm).
 We perform averaging over 5000 replicas (different realizations of random distributions of active and non-active agents).
The maximum number of time steps is taken $t=400$ (typical times for  reaching the
stationary state for the cases studied below are found  in general at $t<100$).

\section{Cluster distribution on disordered lattice} \label{III}

\begin{figure}[b!]
	\begin{center}
		\includegraphics[width=70mm]{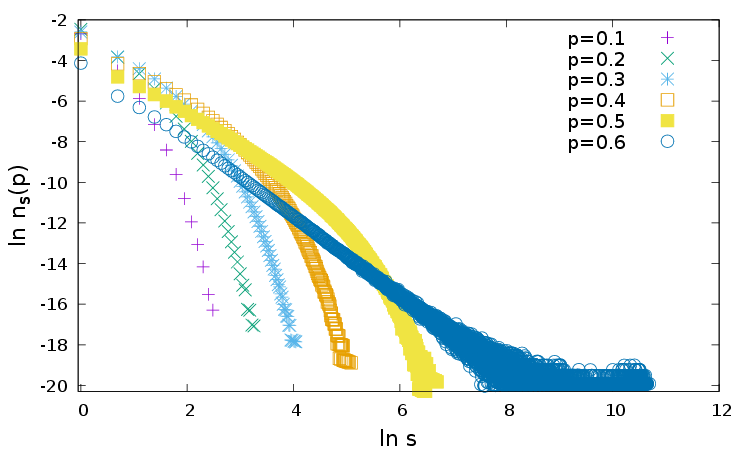}
		\includegraphics[width=70mm]{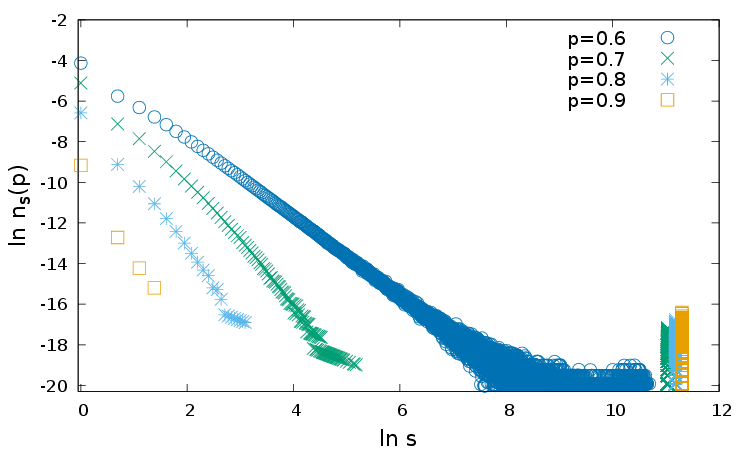}
		\caption{Number of clusters per lattice site  $n_s(p)$ as a function of  cluster size $s$ for a simple square lattice of size $300\times 300$  with different concentrations $p$ of susceptible
			sites in a double logarithmic scale.}
\label{ppure}	
\end{center}
\end{figure}

The problem  under consideration  is closely related to the well known site percolation problem \cite{Staufer}.
Indeed, randomly choosing the fraction $p$ of lattice sites, one can   identify the subgraphs of $s$ linked susceptible individuals  as clusters of size $s$.
The probability $P_s(p)$ to find a cluster containing $s$ susceptible sites is analyzed next.
It is straightforward to obtain the probability to find a single susceptible site  surrounded by immune. As long as these events are independent, this is obtained as
a probability to find a susceptible site $p$ times probabilities of four neighbor sites to be immune, $(1-p)^4$:
\begin{equation}
P_1(p)=p(1-p)^4.\label{p1}
\end{equation}
 In turn
\begin{eqnarray}
&&P_2(p)=4p^2(1-p)^6,\\
&&P_3(p)=3p^3(4(1-p)^7+2(1-p)^8),
\end{eqnarray}
and so on. One has: $P_s(p)=sn_s(p)$ with $n_s(p)$ being the number of $s$-cluster per one site. Exact values for $n_s(p)$ with $s$ up to
17 were obtained in Ref. \cite{Sykes76} for the problem of percolation on a simple square lattice.

In general, one can write
$n_s(p)=p^sD_s$.  In a  low density limit of small values of
$p$ (concentration of susceptible sites well below the percolation threshold $p_c=0.592746$ \cite{Ziff94}),
 only clusters of relatively small size can be found in a system.
At the percolation threshold, when the spanning cluster appears, one obtains the scaling behaviour
\begin{equation}
D_s\sim s^{-\Theta}
\end{equation}
with the Fisher exponent $\Theta =187/91$ \cite{Staufer}. Note that for a finite system, the power law is restricted to small
$s$ only and naturally breaks down for cluster sizes comparable with the size of a system \cite{Chelakkot12}.

{ 
To extract  clusters of different sizes numerically, we apply an algorithm developed by Hoshen and Kopelman \cite{Hoshen}. 
This algorithm is successfully applied in studies of percolation  phenomena in disordered environments \cite{Staufer79,Aharony87,Blavatska08,Willems12, Lapshina19,Kotwica19,Oliveira20} . 
As the first step of the algorithm, all susceptible sites of the lattice are  labeled (numbered in an increasing order). At the second step,
 for each of the labeled sites (say, for the site with the label $n$), we check whether its nearest neighbors are also susceptible. 
  If yes,  two possibilities appear.    If the label of the neighbor is larger than $n$, we change the label of the neighbor to $n$.  If the label of the neighbor is smaller than $n$,  we change the label of the site $n$ to that of the neighbor. This procedure is applied until no more changes of site labels is needed. As a result, we obtain groups of clusters of  susceptible sites of  different sizes, where all the sites in a given group have the same label. 
}
Our numerical data for $n_s(p)$ are given in
Fig. \ref{ppure} in a double logarithmic scale.
On the upper panel of Fig. \ref{ppure} we present results for $p$ values increasing from $p=0.1$ up to the percolation threshold.
Whereas  only small clusters are present in system at small $p$, the probability to find larger clusters increases with increasing $p$.
The  scaling behavior in vicinity of the percolation threshold $p\simeq0.6$ is observed, corresponding  to a straight line in a log-log plot.
One notices the pronounced fluctuations in sizes of  large clusters emerging in a system at the threshold point.  Also, as noted above,
the violation of the scaling law at large $s$ is caused by finite system size.
The lower panel of Fig. \ref{ppure} shows results for $p$ values above the percolation threshold. The probability of observing  small clusters is
decreasing with increasing $p$, whereas the fraction of sites in spanning clusters increases.

\begin{figure}[t!]
	\begin{center}
		\includegraphics[width=80mm]{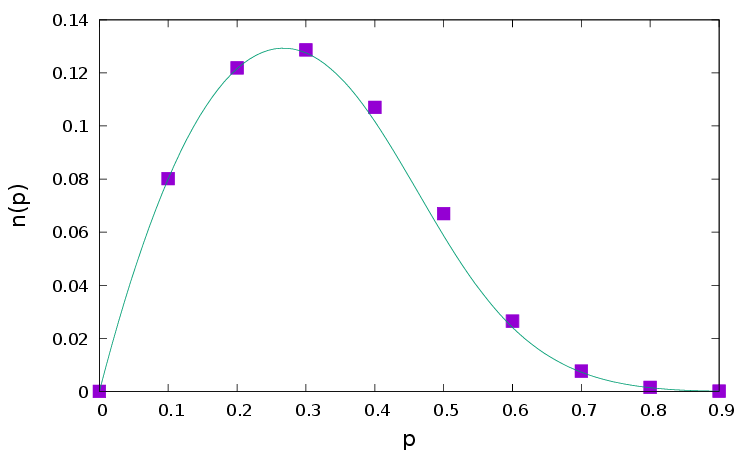}
		\caption{ \label{div}
			Symbols:  our numerical values for  $n(p)$  as a function of $p$.  Lines: analytical results for  $n(p)$  obtained  with making use of
			exact values for $P_s(p)$  given in  Ref.   \cite{Sykes76}. }
	\end{center}
\end{figure}

Let us estimate  also the ``diversity'' of clusters, i.e. the number of all possible clusters, that can be found on a lattice
with given fraction $p$ of susceptible sites.
The lattice of size $L\times L$ contains  $n_s(p)\times L^2$ different clusters of size $s$, and
thus $\sum_s n_s(p) \times L^2$ different
clusters of any size. Effective ``diversity'' of clusters per one site  $n(p)=\sum_s n_s(p) $
can be estimated, using available analytical values for $n_s(p)$ from Ref. \cite{Sykes76} mentioned above. These estimates are compared
with our numerical data in Fig. \ref{div}.
As expected, $n(p)$  increases with $p$, when $p$ values are relatively small, and there is a large amount of small clusters.
At larger $p$, small clusters tend to segregate into larger ones, so the number of different clusters decreases.
Nice coincidence of analytical results that take into account only  small clusters up to $s=17$ and numerical data
is due to the fact,
that though at large $p$ the prevailing number of sites is located in very large clusters, the  number of
such clusters is very small, so that the main contribution into  $n(p)$ is given by numerous  small clusters.

The above distributions will play an important role in further analysis of epidemic processes in such system.
Thus, the considered population  has a ``patchy'' structure and contains small and large groups of susceptible individuals, contacting either
between themselves or with surrounding immune agents. Related patchy system, when infection is allowed to spread only within separated populated areas was
studied recently in Ref. \cite{Athithan14}.

\section{Results for spreading processes} \label{IV}

\begin{figure}[b!]
	\begin{center}
		\includegraphics[width=80mm]{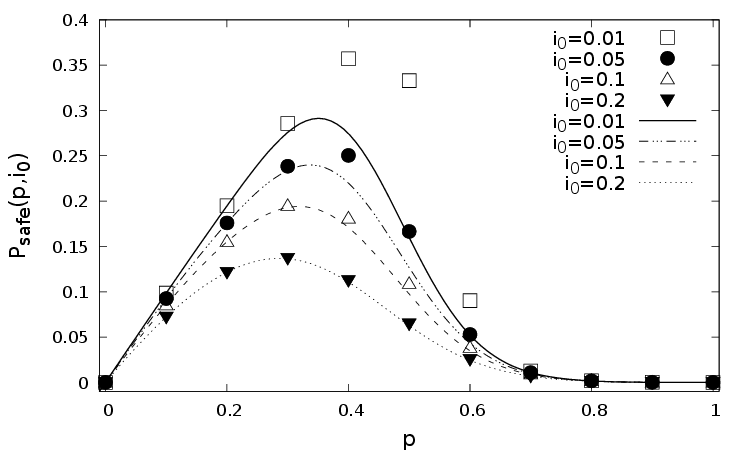}
		\caption{ \label{psi}
			Symbols:  numerical values for  $P_{\rm safe}(p,i_0)$  as function of $p$ at  different $i_0$. Lines: analytical results for
			$P_{\rm safe}(p,i_0)$  obtained on the base of Eq. (\ref{psafe}) with making use of
			exact values for $P_s(p)$  given in  Ref.   \cite{Sykes76}. }
	\end{center}
\end{figure}

We apply the cellular automaton mechanism to study the epidemic process on  {disordered lattices}, described in the previous Section.
We take different values of $p$ in a range $0.1$ to $1.0$ with step $0.1$.
and perform the averaging of all observables of interest over an ensemble of $5 000$ replicas of random realizations of
 {susceptible sites configurations}.

At time $t=0$,  a small fraction $i_0$ of randomly chosen susceptible individuals is supposed to get  infected.
The disease starts to spread from infected to susceptible agents. Taking into account the patchy structure of population under consideration, we
immediately conclude, that
if at $t=0$  in  some cluster of  susceptible agents no one site gets infected, this cluster will not be touched by epidemic process in any way and
remains safe till the epidemic terminates.
 The probability $P_s(p,i_0)$, that cluster of size $s$  is not affected by infection is given by:
\begin{equation}
P_s(p,i_0)=(1-i_0)^sP_s(p). \label{pcon}
\end{equation}
and thus
 \begin{equation}
 P_{{\rm safe}}(p,i_0)=\sum_{s}P_s(p,i_0) \label{psafe}
\end{equation}
gives the total fraction of individuals, which are lucky to be in such safe patterns,
provided that $i_0$ of initial $p$ susceptible sites have been infected.  {The sum in (\ref{psafe}) spans
all clusters of susceptible sites.}

We can get analytical estimation of this value by taking available exact values for $P_s(p)$ up to $s=17$ as given in  Ref.   \cite{Sykes76}
and substituting them into (\ref{pcon}).
 Resulting curves at  several values of $i_0$ are presented  in Fig. \ref{psi}.
 It is important to note the limitation of  {the analytical estimates}, since they do not take into account the contribution of larger clusters.
However, at any $p$ there is a larger amount of  small clusters than the large ones, as discussed in the previous Section.
The ``diversity'' (number of different clusters)  {$n(p)$}  increases with $p$, until it
reaches some critical value, as shown in Fig. \ref{div}. This can help us to
understand the general tendency  of $P_{{\rm safe}}(p,i_0)$  curves.  At fixed $i_0$, number of unaffected clusters grows with increasing $p$, since the number of
different clusters grows, and thus the probability for one of them to get infected gets  lower.
On the other hand, above the threshold value  of $p$ the number of various clusters decreases (small clusters start to segregate into larger structures), which
makes it easier to infect one of them.
Let us compare the analytical estimates with our numerical results for $P_{{\rm safe}}(p,i_0)$,  {as shown in  Fig. \ref{psi}}.
Discrepancy at small $i_0$ is explained by the fact, that also the larger clusters (despite the small amount of them) which are not taken into account in
analytical prediction,
with higher probability remain unaffected by infection.  At larger $i_0$, there is perfect coincidence with analytical prediction. This tells us, that mainly
the clusters of small size remain safe in such processes.

\begin{figure}[t!]
	\begin{center}
		\includegraphics[width=80mm]{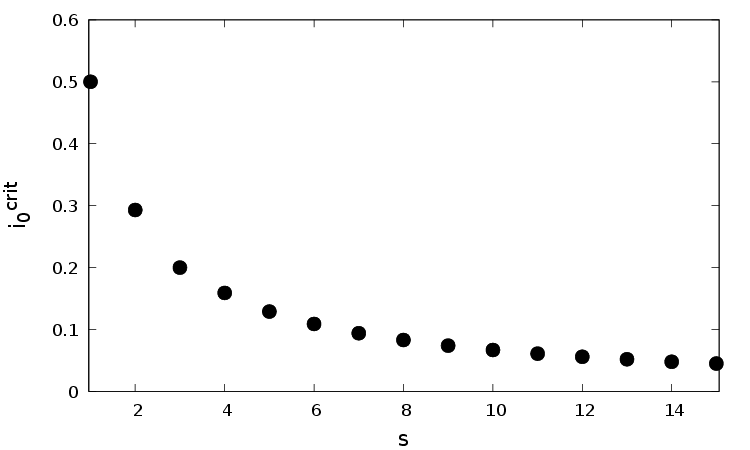}
		\caption{ \label{peq}
			Critical value of the initial ratio of  infected agents $i_0^{\rm crit}$ as a function of cluster size $s$. For given $s$, at $i_0>i_0^{\rm crit}$ the number of
			infected clusters prevails the number of the safe ones.}
	\end{center}
\end{figure}

One can estimate the critical value of $i_0$  for each cluster size $s$, above which there are more infected clusters  than safe at given $p$.
 {This value $i_0^{\rm crit}$ can be obtained} from equality condition for  probabilities to have the  safe and infected clusters
$P_s(p)(1- (1-i_0^{\rm crit})^{s})=P_s(p)((1-i_0^{\rm crit})^{s})$, and is thus
 independent on $p$ and can be easily calculated analytically. The results are presented in Fig. \ref{peq}.

 {Another interesting question is,  whether} there exist a possible minimum value of $i_0$ at which all the existing susceptible patterns
in a system get infected. This is possible in  the case, when  each  cluster of susceptible sites contains at least one infected agent. Thus this minimum
value of $i_0$ is strictly defined by the number of different clusters at given $p$, as presented in Fig. \ref{div}. Note however, that in the case of
random distribution of infected sites
probability of such event  is negligibly small. So in general, introducing any fraction of infected individuals $i_0<1$ into the
patchy lattice one cannot end up with  completely infected system
and  some fraction of individuals  $P_{{\rm safe}}(p,i_0)>0$ will remain locked in the safe patterns.

The above results concern any spreading process on a disordered lattice. Let us now discuss the outcomes of specific spreading scenarios.

\subsection{SI model}

\begin{figure}[t!]
	\begin{center}
		          \includegraphics[width=40mm]{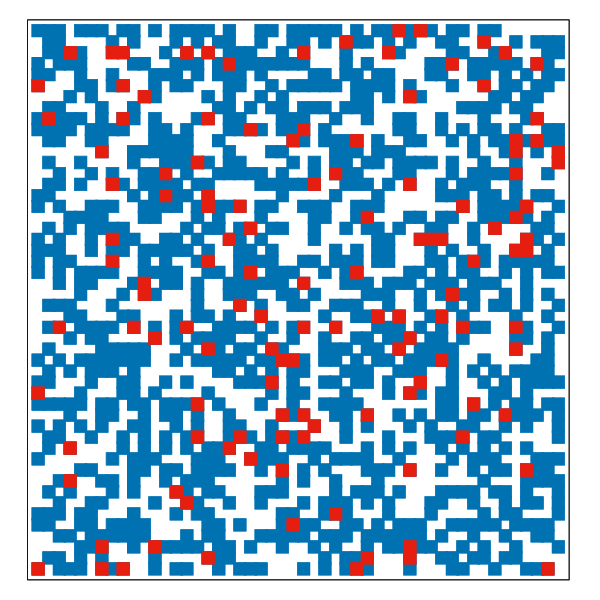}
                     		\includegraphics[width=40mm]{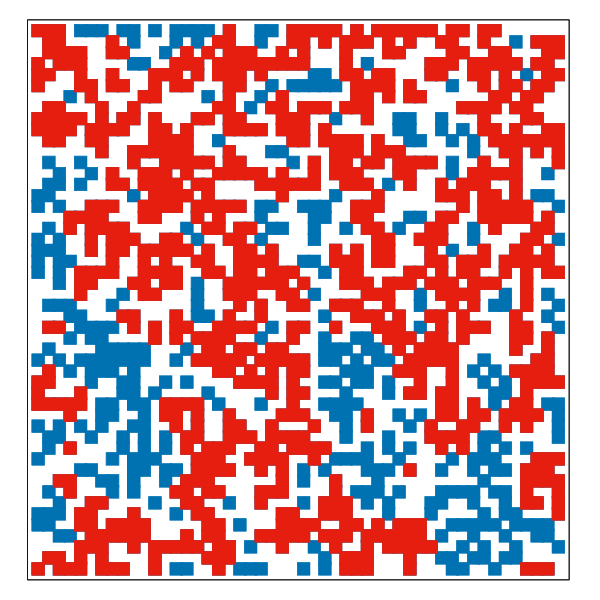}
	\end{center}
\caption{(Color online) Left:   {Square lattice with a randomly distributed fraction} $p=0.5$ of
susceptible agents at initial time $t=0$.  {Share $i_0=0.1$ of the susceptible sites} (shown by red squares) gets infected.
Right: Configuration of the same system after the epidemic process within the  {SI scenario}
reaches equilibrium. Video is available as a supplementary material.}\label{sipicture}
\end{figure}

Within this model, the initially infected agents are spreading infection by contact with their neighbors  within the
clusters, to which they belong, until the equilibrium state is reached.  {Snapshots of configurations of
infected and susceptible agents at the initial time and after reaching the equilibrium are shown in Fig. \ref{sipicture}.
The initial concentration of infected sites of the whole lattice  $I(0)$ is related to  $i_0$ via:
$I(0)=p\times i_0$.}
 We took the value for infection ratio $\beta=0.5$
to make the evolution process quite fast, though the equilibrium values of ratios of infected and susceptible individuals
$I^*(p)$ and $S^*(p)$ do not depend on $\beta$.

\begin{figure}[b!]
	\begin{center}
                   		\includegraphics[width=80mm]{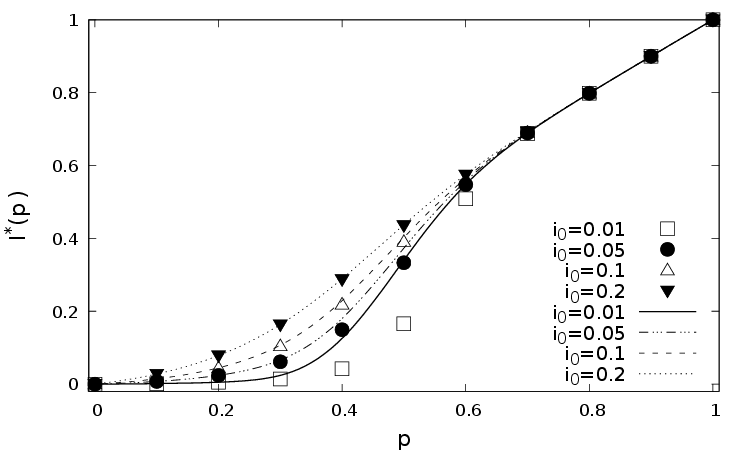}
 		\caption{ \label{siinfl}Symbols:  numerical values for $I^*(p,i_0)$ of SI model as functions of $p$ at  different $i_0$.
 			 Lines: analytical results for  $I^*(p,i_0)$ obtained on the base of Eq. (\ref{inf}) making use of the
exact values for $P_s(p)$  given in  Ref.   \cite{Sykes76}. }
	\end{center}
\end{figure}

Note that the estimates of  $I^*(p,i_0)$ and $S^*(p,i_0)$
can be obtained straightforwardly by making use of conclusions from the previous Subsection.
Indeed, the total fraction $P_{{\rm safe}}(p,i_0)$ of safe patterns, surrounded by immune individuals,
  gives the fraction of susceptible individuals in equilibrium: $S^*(p,i_0)=P_{{\rm safe}}(p,i_0)$
and thus we get:
\begin{equation}
I^*(p,i_0)=p-S^*(p,i_0)=p-P_{{\rm safe}}(p,i_0). \label{inf}
\end{equation}
Again, recalling the definition (\ref{psafe}), making use of  available exact values for $P_s(p)$ up to $s=17$ as given in  Ref.   \cite{Sykes76}
and substituting them into (\ref{inf}), we obtain  analytical estimates
for  $I^*(p,i_0)$,   presented  in Fig. \ref{siinfl} at several values for $i_0$.
These results should be compared with our numerical data for $I^*(p,i_0)$. Again,
 at  larger values of  $i_0$, where mainly the smaller clusters remain untouched by infection,
 there is perfect coincidence with the analytical predictions.

\begin{figure}[b!]
	\begin{center}
		          		\includegraphics[width=40mm]{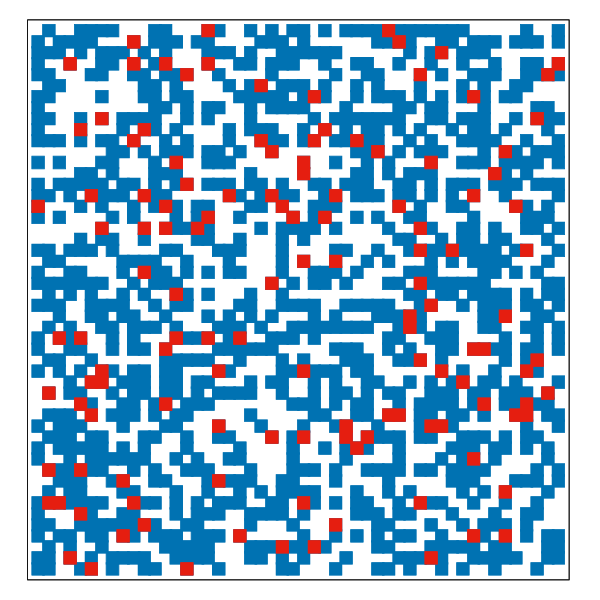}
	          		\includegraphics[width=40mm]{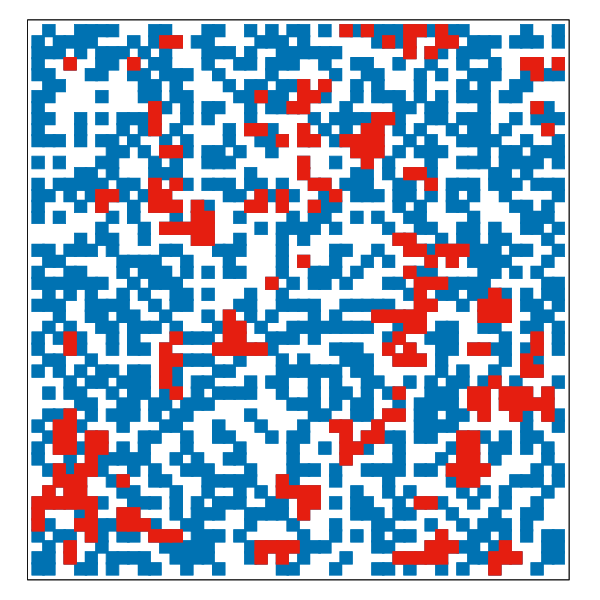}
	\end{center}
\caption{\label{sispicture}(Color online) Left:  {Square lattice with a randomly distributed fraction  $p=0.5$ of susceptible agents at
	initial time $t=0$. Share of $i_0=0.1$ of the susceptible sites (shown by red squares) gets infected.
Right: Configuration of the same system after the epidemic process within the SIS scenario  with the curing rate $\gamma=0.2$  
reaches equilibrium. Video is available as a supplementary material. } }
\end{figure}

\subsection{SIS model}

Within this model, the initially infected individuals  are spreading infection by contact with their neighbors with infection rate
$\beta$ or get cured with curing rate $\gamma=1-\beta$
until the equilibrium state is reached.
 {Snapshots of configurations of
infected and susceptible agents at the initial time and after reaching the equilibrium are shown in Fig. \ref{sispicture} for the same
initial conditions as those of Fig. \ref{sipicture} for the SI scenario. As expected, one notes essentially smaller number of
infected sites in the equilibrium $I^*(p)$.}
An essential feature here is that  any cluster of size $s$, which gets infected initially,
 can become completely ``cured'' in a course of the SIS process and once it happens, all sites in this cluster
remain susceptible.  Thus at the  {endemic} state of SIS process, when the  fraction of infected
sites reaches its stationary value $I^*(p,i_0)$,
one observes in general the larger
amount of patterns of susceptible agents, than in the case of SI process. Indeed, in addition to the set of
``safe'' patterns $P_{{\rm safe}}(p,i_0)$,  which were not affected by the infection from the very beginning
(given by Eq. (\ref{psafe})),  also the set of   ``cured'' patterns $P_{{\rm cured}}(p,i_0)$
 {contributes.}

\begin{figure}[t!]
	\begin{center}
		\includegraphics[width=80mm]{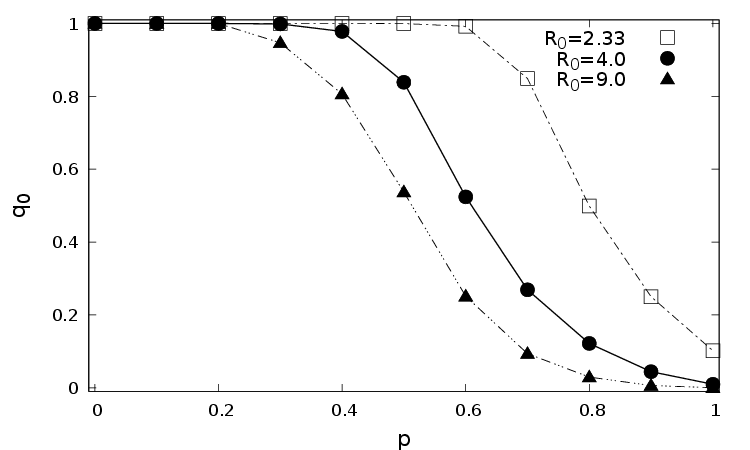}
		\caption{ \label{p0}  Averaged probability $q_0$ to have no infected neighbors for a given lattice site as a function of $p$
			at $i_0=0.1$  and curing rate $\gamma=0.1, 0.2, 0.3$  from  bottom to top (corresponding $R_0$ values are given in the legend).}
	\end{center}
\end{figure}

In the case of disease with repeated infection the only condition for a given individual to stay healthy (once it is cured) is the absence of
infected individuals in its neighborhood. Thus, the probability for cluster of size $s$ to become completely ``cured" is proportional to $q_0^s$,
where $q_0$ is the probability, that in the stationary state no one of four nearest neighbors of a given susceptible site  is infected.
 {The value of}
$q_0$ is easily  estimated numerically as the averaged number of susceptible nearest neighbors of a given susceptible agent;
 {corresponding numbers are} presented in  Fig. \ref{p0} at $i_0=0.1$.
As expected, the higher is the curing rate $\gamma$, the larger is the probability for a given site to be surrounded by only non-infectious neighbors.
 Note that these values are dependent also on the initial concentration $i_0$  {slightly decreasing with an increase of $i_0$.}

\begin{figure}[b!]
	\begin{center}
		\includegraphics[width=80mm]{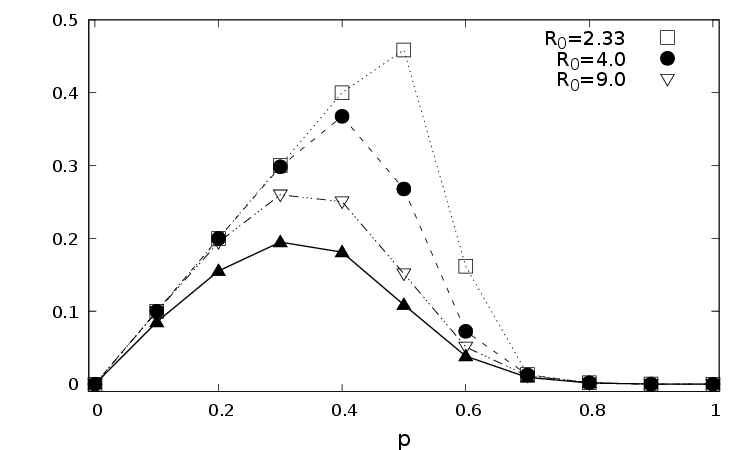}
		\caption{ \label{psis}
			Lower curve: Fraction of susceptible sites $ P_{\rm safe}(p,i_0)$ at $i_0=0.1$ as a function of $p$ (restoring the corresponding
			curve from Fig. \ref{psi}).
		Other curves:  Fractions of susceptible sites $P_{SIS}(p,i_0,\gamma)$  which can be found in a lattice after  the SIS process at
		$i_0=0.1$ and  $\gamma=0.1, 0.2, 0.3$  (from  bottom to top) reaches equilibrium. Lines are guide to the eye.  }
	\end{center}
\end{figure}

Thus, the total fraction of susceptible clusters which can be found in a system after the equilibration of SIS process, 	
is given by:
\begin{equation}
P_{SIS}(p,i_0,\gamma)=P_{{\rm safe}}(p,i_0)+P_{{\rm cured}}(p,i_0,  {\gamma}).
\end{equation}
The second term in this equation
 contains the fraction of individuals in clusters,  which were originally infected but completely  cured  in SIS process.
 In Fig. \ref{psis} we give the estimates based on our numerical data.

\begin{figure}[b!]
	\begin{center}
		\includegraphics[width=80mm]{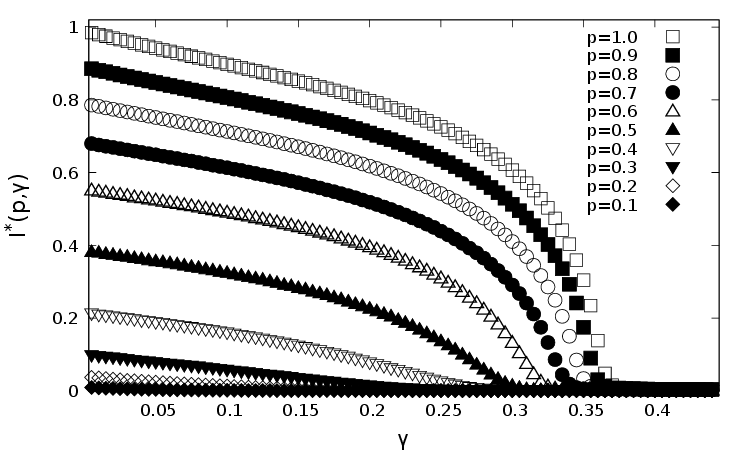}
		\caption{ \label{sisphase} Equilibrium values of $I^*(p,\gamma)$ of SIS model  at $i_0=0.1$ and various $p$ as functions of curing rate $\gamma$.}
	\end{center}
\end{figure}

\begin{figure}[t!]
 	\begin{center}
 		\includegraphics[width=80mm]{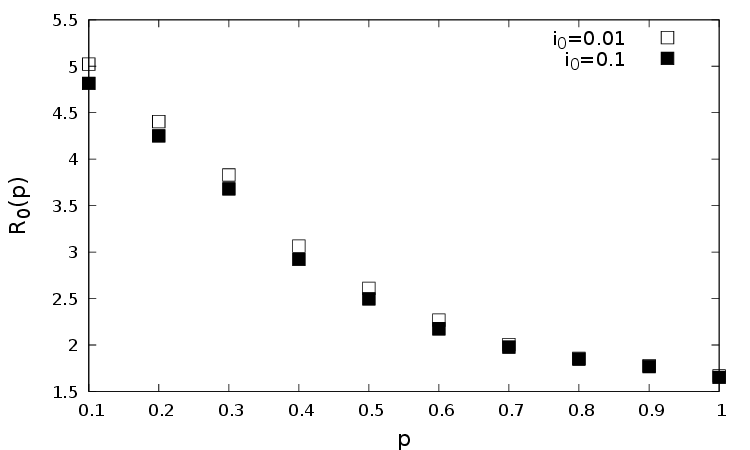}
 		\caption{ \label{R0p}  {The threshold value of the reproductive number}  $R_0^c(p)$ as a function of $p$  {for two  initial
 		concentrations of infected sites}  $i_0$}
 	\end{center}
 \end{figure}

In a course of time evolution, the ratio of infected individuals $I(t)$ reaches the equilibrium value  $I^*(p,\gamma)$.
Corresponding numerical data  are given in Fig. \ref{sisphase}. As expected,  {for fixed  $\gamma$ the final equilibrium fraction} of
infected individuals decreases with the decrease of $p$. The data shown in Fig. \ref{sisphase} demonstrate a continuous transition from the state
when $I^*\simeq 0$ (no endemic occurs) and $I^*>0$ at certain critical value of
curing rate $\gamma_c$. To estimate $\gamma_c$ at various $p$, we apply the fitting to the power form
\begin{equation}
I^*=A(\gamma_c-\gamma)^{\beta'}
\end{equation}
where $\beta'$ is the critical exponent for the order parameter with a mean-field value  {$\beta' = 0.5$} \cite{crit}.
In turn, we can estimate the threshold values of reproduction number according to $R_0^c=(1-\gamma_c)/\gamma_c$.
 Results of fitting of our data are given in Fig. \ref{R0p}.

At $p=1$, our estimates are in  agreement with $R_0^c=1.646$ obtained in \cite{Ilnytsky16} as well as the estimates of $\lambda_c=1.64872(3)$ of contact
processes \cite{Sabag}.
Note that at $p<1$ our results could have been compared with those obtained for contact processes on disordered lattices where only fraction $p$ of sites
are available for spreading
\cite{Moreira96}. However, two different types of ensemble averaging are performed in our work and in Ref. \cite{Moreira96}. Whereas we put the fraction $i_0 $
of initially infected sites randomly and apply averaging over all clusters of initially susceptible sites in the systems, in the
mentioned study only averaging over the spanning percolation cluster was performed and the  process started with single active site on this cluster. As a result,
the values for  $\lambda_c$ at various $p$ obtained in that study are larger than our corresponding $R_0^c(p)$.
Note also, that at $p<1$ the results for $R_0^c$ are sensitive to the value of initial concentration $i_0$.  Plots of Fig. \ref{sisphase} reflect behavior of a
finite-size system and therefore manifest both universal and non-universal features.
	Indeed, for large values of $p>p_c\simeq 0.592746$ \cite{Ziff94} there is a non-zero probability to find a spanning cluster of active sites.
	Therefore, an infinite system possesses at $\gamma=\gamma_c(p)$ a well-defined transition to an endemic state  $I^*(p,\gamma)\neq 0$  for any non-zero
	value of $i_0$. This is reflected by upper curves in Fig.   \ref{sisphase}. On opposite, for $p<p_c$ probability to have a macroscopic
	equilibrium value $I^*$ depends of an initial value of $i_0$, as shown in Fig. \ref{R0p}.

\subsection{SIR model}

Within this model, the initially infected individuals  are spreading infection by contact with their neighbors with infection
rate $\beta$ or get removed with curing rate $\gamma=1-\beta$  until the equilibrium state is reached.
 {Typical configurations of
infected and susceptible agents at the beginning and after reaching the equilibrium are shown in Fig. \ref{sirpicture} for the same
initial conditions as those of Figs. \ref{sipicture}, \ref{sispicture} for SI and SIS scenarios, correspondingly.

\begin{figure}[b!]
	\begin{center}
		\includegraphics[width=40mm]{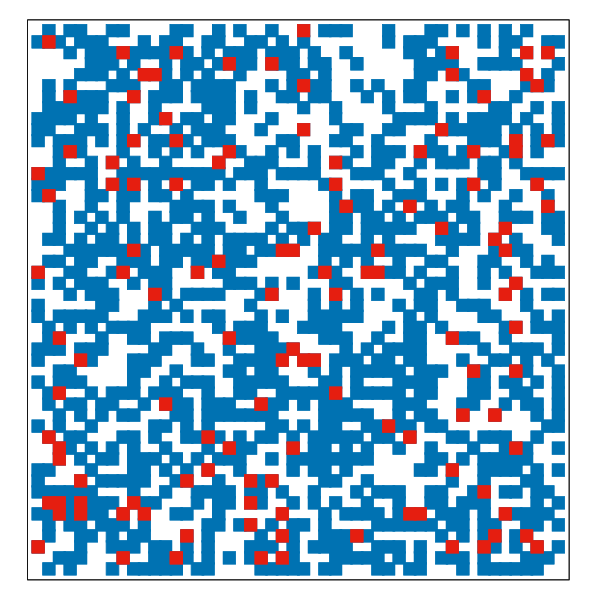}
	          		\includegraphics[width=40mm]{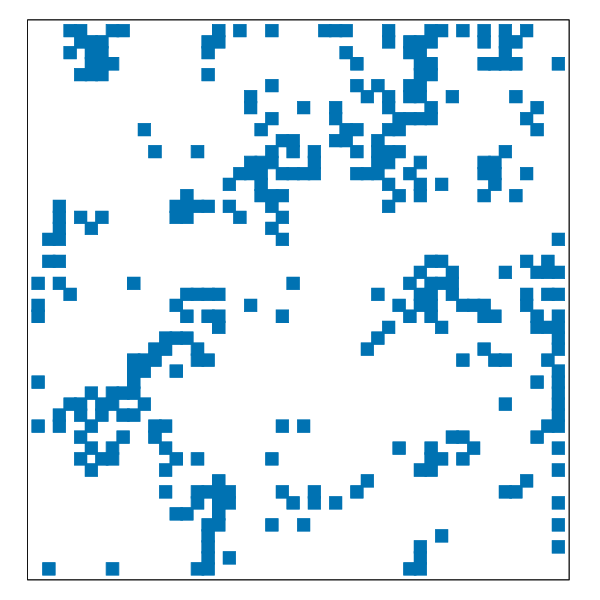}
	\end{center}
\caption{\label{sirpicture} (Color online) Left:  {Square lattice with a randomly distributed fraction  $p=0.5$
of susceptible agents at
	initial time $t=0$. Share of $i_0=0.1$ of the susceptible sites (shown by red squares) gets infected.
Right: Configuration of the same system after the epidemic process within the    SIR scenario  with the curing rate $\gamma=0.2$  reaches equilibrium.}
Video is available as a supplementary material.}
\end{figure}

An obvious difference is that since the}
repeatedly infection is not allowed in course of    SIR process,
it terminates with some equilibrium fraction $R^*(p, {i_0}, \gamma)$ of individuals
 becoming removed (or with the life-long immunity).
The rest of individuals  remain susceptible and thus can be
affected in the case of the  {subsequent} wave of epidemic.  {In the case of    SIR scenario,
the fraction of individuals in susceptible patterns is given by:
\begin{equation}\label{Psir}
 P_{SIR}(p,i_0,\gamma) = S^*(p, i_0,\gamma)=p-R^*(p,i_0,\gamma)\, .
\end{equation}
}

\begin{figure}[t!]
	\begin{center}
            \includegraphics[width=80mm]{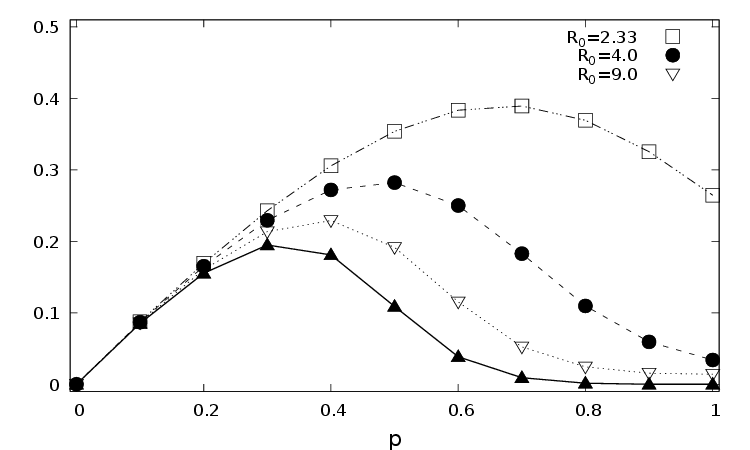}

 \caption{ \label{psir}
Lower curve: Fraction of susceptible sites $ P_{{\rm safe}}(p,i_0)$ at $i_0=0.1$,  as function of $p$.
  Other curves:  Fraction of susceptible sites $P_{SIR}(p,i_0,\gamma)$  which can be found in a lattice after     SIR process at $i_0=0.1$ at  $\gamma=0.1, 0.2, 0.3$
  (from  bottom to top) reaches equilibrium, as functions of $p$. Lines are guide to the eye.  }
	\end{center}
\end{figure}

\begin{figure}[b!]
	\begin{center}
		\includegraphics[width=80mm]{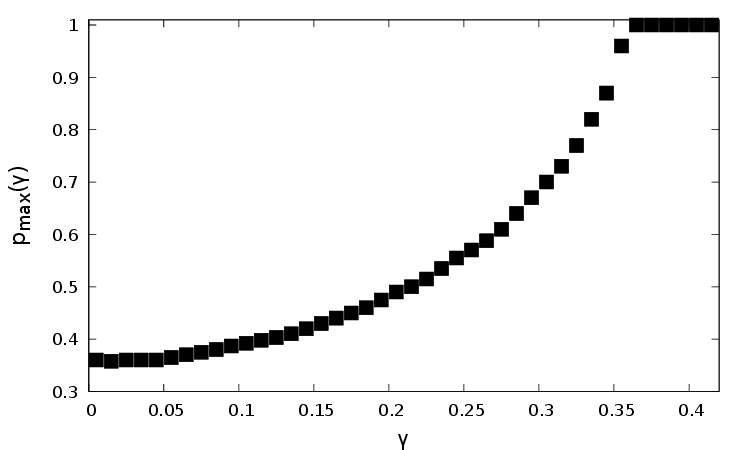}
		\caption{ \label{pmax} The value of $p$, at which a
		fraction of non-affected patterns reaches maximum value,  as a function of curing rate $\gamma$.}
	\end{center}
\end{figure}

Again  we expect, that the total fraction $P_{SIR}(p,i_0,\gamma)$ of susceptible individuals after the termination of    SIR process is
 larger than the fraction in non-affected patterns $P_{\rm safe}(p,i_0)$.  {Indeed, besides $P_{\rm safe}(p,i_0)$ it includes
 also contributions
 from those infected clusters, where the infection spreading has been terminated keeping some agents in the {\em S} state. This is
 further demonstrated by Fig. \ref{psir} where} we present our simulation data for $P_{SIR}(p,i_0,\gamma)$ in comparison with $P_{{\rm safe}}(p,i_0)$.

 Note that
for each value of the curing rate $\gamma$ there exists a corresponding value  $p_{\rm max}(\gamma)$  at which $P_{SIR}(p,i_0,\gamma)$ reaches the maximum.
 After
performing a more refined simulation with $p$ changed in a range $[0 \ldots 1]$ with a step $0.001$ we obtain the data for
 $p_{{\rm max}}(\gamma)$ as a function of $\gamma$, see Fig. \ref{pmax}.  Recalling that $1-p$ can be treated as the fraction of individuals that get
 vaccinated before the spreading process starts, these results can be used for prediction of efficiency of vaccination for disease with given curing
 rate $\gamma$. The larger is the value of $\gamma$ (and thus the weaker is the infection),
 the larger is the critical value $p_{{\rm max}}(\gamma)$  and reversely the smaller is the fraction of agents $1-p_{{\rm max}}(\gamma)$ who need to
 be vaccinated in order to have maximal values of  unaffected individuals in the equilibrium state.  For the case of diseases with sufficiently
 large values of $\gamma$, the infection is so weak that practically no vaccination   is needed: $p_{{\rm max}}(\gamma)$ tends to $1$, therefore
 the number of vaccinated persons, needed to obtain the maximum of unaffected individuals, tends to zero.

\begin{figure}[t!]
	\begin{center}
                   		\includegraphics[width=80mm]{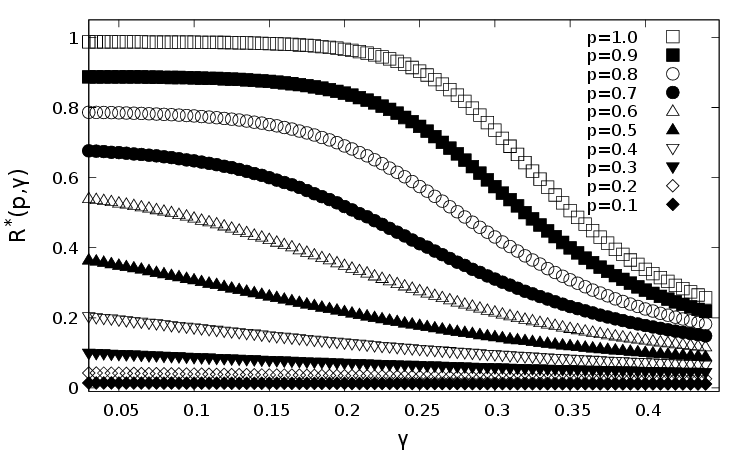}
 		\caption{ \label{sirphase} Equilibrium value of $R^*(p,\gamma)$ of the    SIR model  at $i_0=1$ and various $p$
 		as a function of the curing rate $\gamma$.}
	\end{center}
\end{figure}

Infection spreading throughout the lattice results in an increase of clusters of affected sites (infected and eventually removed).
The curing rate critical value $\gamma_c$ separates a state with no epidemic (corresponding fraction of removed sites $R^*\sim 0$)
and a post-epidemic state with considerably large value of $R^*$.
 {The value of $\gamma_c$ can be estimated by analyzing the removed cluster distributions
from the condition of emergence of the spanning percolation cluster} \cite{Ziff10,Souza10}. In the case when spreading starts with
a single infected site, the process is closely related to standard percolation
growth, when an active (infective) site spreads to susceptible nearest neighbor with the given bond probability,
corresponding in this case to the infection rate $\beta$. This approach however cannot be applied straightforwardly in our problem,
where we put the fraction $i_0 $
of initially infected sites randomly on various clusters of susceptible sites available at $p<1$.

In Fig. \ref{sirphase} we present our results obtained for $R^*(p,\gamma)$ as function of $\gamma$ for various $p$.
The very rough estimates for $\gamma_c$ can be obtained from the condition of pronounced decrease of  $R^*(p,\gamma)$ after the plateau regime.
Let us remind, that at $p=1$ the critical value $\gamma_c=0.1765005(10)$ \cite{Ziff10}, and it is obviously decreasing with the decrease of $p$.
As expected, the smaller is the fraction of susceptible agents in population, the smaller is the critical value of curing rate (and thus
the larger is the corresponding value of  the basic reproduction number $R_0^c$), above
 which the infection spreading process terminates without causing a global epidemic process.

\section{Conclusions} \label{V}


In this paper, we have presented quantitative analysis of  a spreading process in a non-homogeneous environment.
To this  end, we have considered three archetypal spreading models, SI, SIS and SIR on structurally-disordered lattices, when
only a part of lattice sites (agents) is active and takes part in spreading
phenomenon. One of possible interpretations of such a
problem statement is infection spreading in population where only a part of individuals is susceptible to disease.
The rest is non-sensitive to infection, as may happen after the first wave of epidemic in over and a part of population
gets immune. Another interpretation may be that some randomly chosen part of agents gets immune due to the formerly performed
vaccination of a part of population.  In our analysis, we have considered
the case, when  the probability of arbitrary chosen
lattice site to be susceptible is uniquely determined by their concentration $p$. In turn, this leads to emergence
of different size clusters of active susceptible sites.
As it becomes evident from the analysis, spreading process
in such ``patchy'' population is characterized  by a number
of distinctive and sometimes unexpected features. In particular,
absence of connetivity between finite-size clusters of susceptible
sites causes existence of the so-called  ``safety patterns'': if no
infected sites fall into such pattern, its state remains unchanged untill
the stationary state is reached. Detailed analysis of
distribution of the ``safety patterns'' enabled us to determine the
fraction of infected agents in a stationary state $I^*(p)$. Subsequently,
this lead to the estimates for the threshold value of the basic reproduction number
$R_0^c$ as a function of active agents fraction $p$ for different spreading
scenarios.

Within the homogeneous mixing hypothesis \cite{SIAM}, consideration of a part of population as
being susceptible  is trivial and leads to renormalization of an active fraction. However, this is
not the case for the lattice model considered here. In this respect it is instructive to compare
processes of spreading and ordering  on an inhomogeneous lattice. To give an example, inhomogeneities
in lattice structure caused by dilution of a magnet by a non-magnetic component, may cause severe
impact on a magnetic phase transition \cite{crit,disord_magn}. However, for a short-range interaction,
the spontaneous magnetization can be non-zero only if the concentration of magnetic sites is above the percolation
threshold, $p>p_{\rm perc}$. No ordering is expected for $p<p_{\rm perc}$: indeed, in this region
only the finite-size clusters of magnetic sites exist, and their contribution to overall
magnetization vanishes in the thermodynamic limit $N\to \infty$. As we have shown above, this is not
the case for the spreading phenomenon
on disordered lattices: there, the  spreading is non-trivial for any concentration of active sites
$p$, also at $p<p_{\rm perc}$. The reason is that although in this region the contribution from each
finite-size cluster vanishes, their steady-state value may be non-zero also in the limit
$N\to \infty$, depending on the basic reproduction number $R_0^c$.

It is obvious that results obtained in our analysis crucially depend on the
distribution of susceptible sites. Here, we have considered the simplest case, when correlations in
distribution are absent. It may be interesting to re-consider the problem for the case when such
correlations are present, taking them from available real-world data. Returning back to the above discussed
comparison of spreading and ordering on inhomogeneous lattices, this example may find its
counterpart in and impact of long-range-correlated impurities
on magnetic phase transitions \cite{long-range-correlated1} and scaling \cite{long-range-correlated2}.
Another work is progress is to apply similar methodology to analyze spreading
processes on complex networks \cite{Pastor-Satorras15}.

\section*{Acknowledgements}

This work was supported in part by National Research Foundation of Ukraine, project ``Science for Safety of Human and Society'' No. 2020.01/0338 (VB)
and by and by the
National Academy of Sciences of Ukraine, project KPKBK 6541230 (YuH)..
It is our pleasure to acknowledge useful discussions with Jaroslav Ilnytskyi and Mar'jana Krasnytska.

\end{document}